\documentclass[final,3p]{elsarticle}

\usepackage{amssymb}

\journal{New Astronomy Reviews}

\begin{document}

\begin{frontmatter}

\title{Massive Stars in the Galaxies of the Local Group}

\author{Philip Massey}

\address{Lowell Observatory, 1400 W. Mars Hill Road, Flagstaff, AZ 86001, USA; phil.massey@lowell.edu}

\begin{abstract}
The star-forming galaxies of the Local Group act as our laboratories for testing massive star evolutionary models.  In this review,
I briefly summarize what we believe we know about massive star evolution, and the connection between OB stars, Luminous Blue Variables, yellow supergiants, red supergiants, and Wolf-Rayet stars.  The difficulties and recent successes in identifying these
various types of massive stars in the neighboring galaxies of the Local Group will be discussed. 
\end{abstract}


\end{frontmatter}

\section{Introduction}

Massive stars contain many of the extremes of astrophysics. Their high masses result in high luminosities, with
energy outputs on the order of a million times that of the sun.  While on the main-sequence as O-type stars, they have nearly the highest effective temperatures of any stars, with only white dwarfs being hotter.   As Luminous Blue Variables (LBVs), they show spectacular outbursts, throwing off large amount of material and brightening by several magnitudes visually.  As yellow supergiants (YSGs), they are among the visually brightest in any galaxy, although, ironically, their identification is often complicated by the plethora of foreground Galactic yellow dwarfs. As cool red supergiants (RSGs), they flirt with the limits of hydrostatic equilibrium, and are physically the largest stars: if you place one of the biggest at the center of the solar system, its photosphere would end somewhere between the orbits of Jupiter and Saturn. As Wolf-Rayets (WRs) they are little more than stripped stellar cores with such strong stellar winds that they would lose ten times the mass of the sun in a mere million years, were they to live that long. Finally, massive stars end their lives as spectacular supernova, briefly outshining the whole of their host galaxies.

\subsection{Motivation}

There are two fundamental reasons for studying massive stars amongst the star-forming galaxies of the Local Group.  First,
these galaxies allow us to perform tests of stellar evolutionary theory in a laboratory where primarily a single variable
can be changed, namely the chemical composition of the stars.
The physical properties of a massive star will be essentially identical at birth regardless
of small changes in the composition.  (Recall that the heavier elements represent only a few percent of the overall composition
of stars.)  But, these trace elements have a large impact on the subsequent evolution of massive stars.   On the main-sequence,
a massive star's high effective temperature and luminosity result in strong stellar winds, which are driven
by radiation pressure on these highly ionized metal atoms, resulting in 
significant mass loss.  A 100$M_\odot$ star could lose half of its mass during its short lifetime!  The
importance of this mass-loss on mass star evolution was first demonstrated by the early models that attempted to include the effects of mass loss (e.g., de Loore et al.\ 1977, 1978;  Chiosi et al.\ 1978, 1979;  Brunish \& Truran 1982).  Modern models continue to demonstrate the importance of mass loss to
massive star evolution (Meynet \& Maeder 2005).   Because these winds are driven by radiation pressure acting on highly ionized
metal atoms, the mass-loss rates $\dot{M}$ depend upon a star's initial metallicity $z$, as something like  $\dot{M}\sim z^{0.7}$ (Vink et al.\ 2001).  Thus, over the $\sim 20\times$ metallicity difference between WLM ($z/z_\odot \sim 0.1$) to M31 ($z/z_\odot \sim 2$),
we expect the main-sequence mass-loss rates to vary by nearly an order of magnitude.  Such differences should be (and are)
reflected in the relative number of evolved massive stars of various kinds, such as the relative number of Wolf-Rayet stars to red supergiants, or the relative number of WC- and WN-type WRs.

Secondly, these massive stars affect the overall evolution and properties of the galaxies themselves, through 
three ``feedback" mechanisms (see, e.g., Oey \& Clarke 2009).    First, their ultraviolet radiation heats dust,
powering the far-IR luminosities of galaxies, while at the same time providing the ionizing radiation that causes
the HII regions (see, e.g., Maeder \& Conti 1994).  
Of course, it is these HII regions which delineate the arms in spiral galaxies, and otherwise reveal
where most of the star formation action is occurring  in irregular galaxies.   Secondly, their strong stellar winds provide significant mechanical
energy input into the interstellar medium, as does their eventual disruption as supernovae (Abbott 1982), shock-heating the
gas to $>10^6$ K (Oey \& Clarke 2009).   This mechanical energy feedback is responsible for the creation of superbubbles
(Pikel'ner 1968; Weaver et al.\ 1977; see discussion in Oey et al.\ 2001).
And thirdly, they are responsible for much of the chemical enrichment of galaxies, particularly of the ``lighter" elements (atomic weight
less than 30), such as carbon, nitrogen, and oxygen (Maeder 1981, Sparke \& Gallagher 2000).   During their red supergiant
phase, massive stars also make a significant contribution to the dust content, particularly for star-burst systems and galaxies
at large look-back times where AGBs have not yet formed (Massey et al.\ 2005a).

In this review, we will first provide an introduction to massive stars, and then follow this with a more detailed description of the current state of our knowledge (and lack thereof) of the content of these stars in Local Group galaxies. 
A much shorter  review on this subject was given by Massey (2010), and an older, but more detailed look, can be found in Massey (2003).  

When we talk about the ``star-forming galaxies of the Local Group," we are referring to the galaxies listed in Table~\ref{tab:galaxies},
an updated version of Table 1 from Massey (2003) and Massey et al.\ (2007b).

\begin{table}
\caption{\label{tab:galaxies}Star-forming Galaxies of the Local Group$^a$}
\begin{center}
\begin{tabular}{l c c c c c c c}
\hline
Galaxy
&Type
& $l$
& $b$
& $M_V$
&$E(B-V)^b$
&Dist (Mpc)  
&log O/H+12 \\ \hline
MW & S(B)bc I-II: & --- & --- & -20.9? & --- & --- & 8.7$^c$  \\
M31 & Sb I-II & 121.2 & -21.6 & -21.2 & 0.13 & 0.76 & 8.9-9.0$^d$ \\
M33 & Sc II-III & 133.6 & -31.3 & -18.9 & 0.12 & 0.83 & 8.3-8.9$^e$ \\
LMC & Ir III-IV & 280.2 & -33.3 & -18.5 & 0.13 & 0.050 & 8.4 \\
SMC & Ir IV-V & 302.8 & -44.3 & -17.1 & 0.09 & 0.059 & 8.0 \\
NGC 6822 & Ir IV-V & 25.3 & -18.4 & -16.0 & 0.25 & 0.50 & 8.1 \\
IC 10 & Ir IV: & 119.0 & -3.3 & -16.3 & 0.81 & 0.66 & 8.2 \\
IC 1613 & Ir V & 129.7 & -60.6 & -15.3 & 0.72 & 0.05$^f$ & 7.9 \\
WLM & Ir IV-V & 75.9 & -73.6 & -14.4 & 0.07 & 0.93 & 7.7 \\
Pegasus & Ir V & 94.8 & -43.6 & -12.3 & 0.15 & 0.76 & 7.9\\
Phoenix & dIr/dSph & 272.2 & -69.0 & -9.8 & 0.15 & 0.40 & ---\\
\hline
\multicolumn{8}{l}{$^a$Data from van den Bergh (2000) and references therein, except as noted.}\\
\multicolumn{8}{l}{$^b$From Massey et al.\ (2007b), except as noted.} \\
\multicolumn{8}{l}{$^c$In the solar neighborhood.} \\
\multicolumn{8}{l}{$^d$From Sanders et al. (2012) at R=12 kpc.}\\
\multicolumn{8}{l}{$^e$From Magrini et al.\ (2007) but see Bresolin (2011) and discussion in Neugent \& Massey (2011).} \\
\multicolumn{8}{l}{$^f$From Sandage (1971).}\\
\end{tabular}

\end{center}
\end{table}

\subsection{A Massive Star Primer}
\label{Sec-primer}

Before delving into the primer, it is worth emphasizing that our understanding of massive star properties and evolution
derives not from observations or theory alone, but from the combination of the two: even something as basic as a star's
luminosity relies not only upon the visual brightness, but also a knowledge of stellar atmospheres in order to relate what
is observed to the meaningful physical quantity.
Within this review, a lot of the emphasis will be given to the observational side, but it would be horribly remiss not
to stress the importance of theory as well.  Just as large telescopes and more efficient detectors have made possible observations only dreamed of in the 1960s, so has the improvement in the physics and computational tools improved both the study of stellar atmospheres and stellar evolution.  Further, just as observers often go to pains to make their data freely available, our theoretician colleagues have gone to similar efforts to share the results of their modeling and codes.   The first non-LTE atmosphere modeling
that included only H and He done by Auer \& Mihalas (1972)  have now evolved to very complex non-LTE codes that include
the effects of line-blanketing and stellar winds, such as CMFGEN (Hillier \& Miller 1998), a code which is made publicly available by the kindness of D. John Hillier.   Similarly, the early evolutionary models that included the effects of mass loss have now evolved
to very sophisticated models that include the effects of rotation and mass loss throughout the H-R diagram (e.g., Maeder \& Meynet 2000a, 2000b; Meyent \& Maeder 2000, 2003, 2005; Ekstr\"{o}m et al.\ 2012; Georgy et al.\ 2012).  The Geneva group has, in
particular, always been very generous by making their models publicly available, encouraging direct comparisons (and sometimes
confrontations) between predictions and observations.

\subsubsection{The main-sequence}

\paragraph{Evolution}

\begin{figure}
\begin{centering}
\includegraphics[width=10cm]{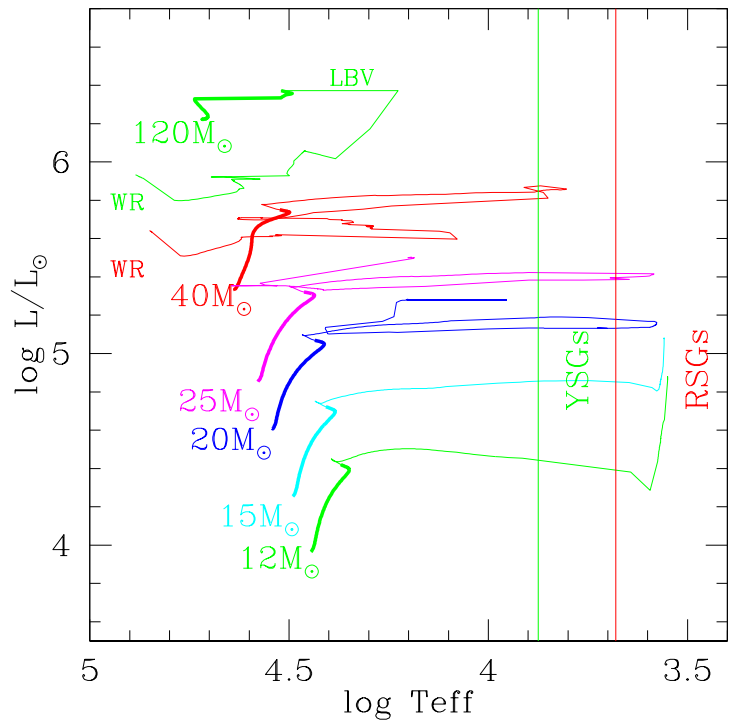}
\caption{\label{fig:HRD} Evolution of massive stars.  The high mass Geneva evolutionary tracks (Ekstr\"{o}m et al.\ 2012) for solar metallicity (z=0.014) are shown ; these models assume an initial rotation velocity of 40\% of the critical break-up speed.
 The initial masses label the beginning part of each track; for simplicity we have not included the 85$M_\odot$ or 60$M_\odot$ tracks.
 The bold sections of the tracks show the main-sequence.
 The general regions of the yellow supergiants and red supergiants  are shown.}
\end{centering}
\end{figure}

Massive stars ($m>8M_\odot$) begin their lives on the main-sequence as O- and early B-type dwarfs.  As these stars age,
they burn hydrogen in their cores via the CNO cycle. The main-sequence (hydrogen-burning) phase lasts for between
3~Myr (120~$M_\odot$) and 30~Myr ($9~M_\odot$), according to the latest solar-metallicity Geneva models (Ekstr\"{o}m et al.\ 2012).  We summarize some of their main-sequence properties in Table~\ref{tab:props}, where we have assigned
spectral types and luminosity classes based upon their evolutionary effective temperatures and surface gravities.
There are several things of note:
\begin{itemize}
\item During the main-sequence these stars change very little in luminosity, with typically a change of only 0.2-0.4~dex.  Past the main-sequence evolution in the theoretical H-R diagram, evolution is pretty much horizontal.  We illustrate this in Figure~\ref{fig:HRD}, using
the latest Geneva evolutionary tracks (Ekstr\"{o}m et al.\ 2012).
\item These stars become {\it supergiants} even before core He-burning begins, i.e., while still on the main-sequence.  To most stellar spectroscopists, that statement would appear to be a tautological impossibility. 
\item Unlike stars of lower mass, where the spectral type is directly indicative of a star's mass,  the spectral type on the main
sequence is more of a phase through which stars with a range of masses may pass.  Thus, it makes no sense to ask, ``What is
the mass of O4~V stars?"    Any star whose evolutionary track causes it to pass through the effective temperature range 42,500-45,500 with $\log g=4.2-3.9$ would be called spectroscopically an O4~V.  This includes any star with a mass
from $\sim$35$M_\odot$ to $\sim$85$M_\odot$.
\end{itemize}

 \begin{table}
 \tiny
\caption{\label{tab:props}Properties of Solar-metallicity Main-Sequence Massive Stars$^a$}
\begin{center}
\begin{tabular}{l c c c c l c c c c l c c c c l}
\hline
 
 &
 & \multicolumn{4}{c}{ZAMS}
 &
 & \multicolumn{4}{c}{Turn-off}
 &
 &\multicolumn{4}{c}{Start of He-burning} \\ \cline{3-6} \cline{8-11} \cline{13-16}
 Mass  
 &$t_H$
 &$T_{\rm eff}$
 &Log   
 &Log g 
 &Spect.
 &
 &$T_{\rm eff}$
 &Log   
 &Log g 
 &Spect.
 &
&$T_{\rm eff}$
 &Log   
 &Log g 
 &Spect. \\
  $M_\odot$
 &(Myr)
 & (K) 
 & $L/L_\odot$
 & [cgs] 
 & Type 
 &
  & (K) 
 & $L/L_\odot$
 & [cgs] 
 & Type 
 &
  & (K) 
 & $L/L_\odot$
 & [cgs] 
 & Type \\ \hline
120 & 3.2  & 52,000 & 6.2 & 4.1 & O2-3~V && 50,500 & 6.2 & 4.0 &O2-3~V&& 29,400 & 6.0 & 2.8 & B0~I \\
  85 & 3.7  & 50,000 & 6.0 & 4.2 & O2-3~V && 48,000 & 6.0 & 4.0 &O2-3~V&& 21,300 & 6.3 & 2.1 & B1~I\\
  60 & 4.4  & 47,000 & 5.7 & 4.2 & O3~V    &&  45,000& 5.8 & 4.0 & O4~V &&   7,100 & 6.1 & 0.2& F5~I \\
  40 & 5.7  & 43,000 & 5.3 & 4.2 & O4~V    && 31,500 & 5.7 & 3.2 & O9~I  &&   6,400 & 5.9 & 0.2 & F8~I\\
  25 & 7.9  & 38,000 & 4.9 & 4.2 & O6~V    && 27,300 & 5.3 & 3.2 & B0.2~I &&   7,900 & 5.4 & 0.9 & F0~I\\
  20 & 9.5  & 35,000 & 4.6 & 4.3 & O7.5~V && 25,800 & 5.0 & 3.3 & B0.5~I && 12,300 & 5.1 & 1.9 & B6~I \\
  15 & 13.5& 31,000 & 4.3 & 4.3 & O9.5~V && 24,300 & 4.7 & 3.4 & B0.5~I &&   9,500 & 4.8 & 1.6 & A0~I\\
  12 & 18.4& 27,900 & 4.0 & 4.3 & B0.5~V && 22,300 & 4.4 & 3.5 & B0.5~I &&   8,200 & 4.5 & 1.7 & A8~I \\
    9 & 31.2& 24,000 & 3.6 & 4.3 & B1~V    && 19,600 & 4.0 & 3.5 & B1.5~III &&   3,800 & 4.1 & 0.6 &M0~I  \\
\hline
\multicolumn{16}{l}{$^a$Based on Ekstr\"{o}m et al.\ (2012), Massey et al.\ (2005b), and Levesque et al.\ (2007).}
\end{tabular}
\end{center}
\end{table}

There are a few other caveats that need to be kept in mind when talking about the main-sequence evolution of massive stars.  
First, the luminosity class (``I" vs ``V", say) of O-type stars depends primarily on lines that are not so much sensitive to 
surface gravity but rather to stellar wind strengths.  The higher the bolometric luminosity at a given
metallicity, the stronger the stellar wind will be ($\dot M\sim L^{2.1}$, according to Vink et al.\ 2000, for $T_{\rm eff}>27,500$~K), 
as there is only a weak dependence upon the effective temperature
in the mass-loss rates for stars this hot\footnote{The mass-loss rates change by 0.01~dex due to temperature from 40,000 to 50,000 K,
according to Vink et al.\ (2000), equation 12.}.    So, stars in the high mass and luminosity range will have the strongest mass-loss rates.  
The luminosity class is primarily based upon the morphology of the He II $\lambda 4686$ line.  This line is both sensitive to mass-loss rates {\it and} effective temperatures, and (for a given spectral subtype) a He~II~$\lambda 4686$ line that is strongly in absorption leads
to a luminosity class ``V", while one in emission leads to a luminosity class ``I".  When He~II~$\lambda 4686$ is in emission, along
with the neighboring N~III $\lambda 4634, 42$ line, we call the O star an ``Of-type", with the implication that the star is of luminosity
class ``I".  (The actual system is a bit more complicated than this; see, e.g., Sota et al.\ 2011.)  The NIII emission behavior has finally
been demonstrated to be dependent on stellar wind strengths, although nitrogen abundance also plays an important role
(Rivero Gonz\'{a}ez et al.\ 2011).  Of course, a significant problem occurs when one looks at O stars in other galaxies that
are more metal-poor than the Milky Way where the various spectral standards have been defined.  Even a high-luminosity O star
in the SMC may fail to show He II $\lambda 4686$ and be labeled a ``giant" or even a ``dwarf", when its absolute visual magnitude
demonstrates it's a supergiant; see discussion in Massey et al.\ (2005b).

The other thing to keep in mind about the evolutionary models is that they are static, not hydrodynamic.  Note that in Table~\ref{tab:props}
I assigned a spectral type of B0~I to the 120$M_\odot$ star at the start of He-burning, but
the situation may be a little more complicated than that.  These stars will have such high mass-loss rates that it is possible, even likely, that the 120$M_\odot$ star will not be identified as an O3-4~V or B0~I, but rather as a hydrogen-rich, WN Wolf-Rayet star, similar to the highest mass (but unevolved) stars seen in the R136 cluster in the LMC and NGC~3603 in
the Milky Way: ``Of-type stars on steroids" (Crowther et al.\ 1995, Massey \& Hunter 1998).

\paragraph{Stellar Winds}

One of the areas of active research is the issue of what the mass-loss rates are during the hot main-sequence stage.
Historically, there have been three methods for measuring the mass-loss rates  in O-type stars: 
\begin{enumerate}
\item P Cygni profile UV resonance lines (e.g., Lamers et al.\ 1987,  Howarth \& Prinja 1989, Haser 1995).  These require
one to know the ionization/excitation fraction and abundance of the element; i.e. at best one measures $\dot{M} \times q$, where
$q$ is the ion fraction of the species.  But even if one knows $q$ to high accuracy, mostly these lines are saturated, leading
to only lower limits on $\dot{M}$. 
\item Recombination lines, such as the H$\alpha$ emission profile, superposed upon photospheric absorption (Klein \& Castor 1978, Leitherer 1988, Lamers \& Leitherer 1993).  This method has often been considered to be more accurate than (1), leading to values that are 10\% or better.  But, one still needs to get everything else right including the velocity law, especially if the winds are weak
and hidden in photospheric H$\alpha$.  In those cases the method is probably a factor of 2 uncertain.
\item IR, (sub)millimeter, and radio continua (Wright \& Barlow 1975, Panagia \& Felli 1975).   Using this method, one compares
the measured flux to that predicted by photospheric models.  Most of the excess is free-free (Bremsstrahlung) and bound-free emission in the stellar winds.  This method has the advantage of being very simple analytically, as long as the winds are assumed to be homogeneous and spherically symmetric.
\end{enumerate}
Both the H$\alpha$ and IR/radio continua methods rely upon the interaction of two particles, and they are therefore referred to as ``density-squared" ($\rho^2$) processes.  In contrast, P Cygni lines are due to scattering; i.e., a photon is absorbed and immediately re-emitted.
In general, the results from the two $\rho^2$ processes agreed, but with the occasional pathological cases (Lamers \& Leitherer 1993, Puls et al.\ 1996).  This happy state of affairs was changed when Fullerton et al.\ (2006) used the FUSE satellite
to measure the mass-loss rates from the P V $\lambda \lambda 118, 28$ resonance doublet.  This side stepped the saturation problem with the P Cygni profiles: since phosphorus is rare, the line isn't saturated.  They found mass-loss rates which were, at first blush,
a factor of 10 lower than what had been assumed previously with the assumption of homogeneous, spherically symmetric winds.  Instead, it now accepted that stellar winds are ``clumped", and not as homogeneous as had been assumed for convenience. As Fullerton et al.\ (2008) wrote, ``Thus, a fundamental consequence of clumping in hot-star winds is that the values of $\dot{M}$ ($\rho^2$)
{\it must be reduced.}  The only question is: by how much?"  Naively, all other things being equal, the answer would be a factor of ten
but current thinking now is that generally the value is more likely a factor of 2 or 3 (Hirschi 2008, Puls et al.\ 2008).  But this remains one of the great uncertainties that motivates observational comparisons of the relative number of massive stars in various stages versus
what is predicted by the evolutionary models with some assumed mass-loss rate.

Knowing the number of progenitor, main-sequence stars relative to the number of their evolved descendants provides a key observational test of stellar evolutionary models, but as we will see in Section~\ref{Sec-MS}, quantifying this is among the most
difficult observational challenges, as the main-sequence phase is when massive stars are the visually faintest.  

\subsection{Luminous Blue Variables}

What happens after the main-sequence, depends upon the mass of the star.  For the highest mass stars, they encounter a little difficulty as they edge to cooler temperatures: the opacities of the lines increase as the temperatures decrease, and soon radiation pressure
equals or exceeds the force of gravity (Lamers 1997).   This point is probably associated with the phase known as the classic
``Luminous Blue Variables," previously known as Hubble-Sandage variables (Hubble \& Sandage 1953) or ``S Doradus" variables.
Of all the various phases through which a massive star passes, the details and implications of the LBV phase are probably the most
poorly understood.  What is generally agreed upon is that the ``classic" LBVs, such as S Doradus and P Cygni, undergo giant ``eruptions" every few hundred years, in which there are large photometric changes (several magnitudes) accompanied by large
amounts of material coming off of the star (see, e.g., Bohannan 1997 and Conti 1997).

The star $\eta$ Car is often cited as the prototypical LBV.  It certainly has been the most heavily studied both ground-based and space-based (e.g., {\it HST}): the SAO/NASA Astrophysics Data System reports nearly 500 refereed papers containing $\eta$ Car in its title,
and over 1200 if conference proceedings are counted as well.  Beyond question it is an interesting object in its own right, but $\eta$ Car is probably not much of a rosetta stone for LBVs in general, as it consists of a 5.5 year binary.  The stars are in an eccentric orbit,
and some have argued that the major eruptions have been triggered by mass transfer at periastron passage (see, e.g., Kashi \& Soker 2010).  The discovery of the binary nature of the system has been relatively recent (van Genderen et al.\ 1994, Damineli et al.\ 1997, Sonneborn et al.\ 2005), and has led to a series of observing campaigns at each new close passage.

Smith \& Owocki (2006) suggest that, if the main-sequence mass-loss rates are actually $\sim10\times$ lower than they the unclumped models indicate (rather than the 2-3$\times$ Puls et al.\ 2008 and others have argued must be the case), the mass loss during the LBV phase may be critically important to the further evolution of massive stars.  
Smith et al.\ (2004) argue for the existence
of ``lower mass" LBVs, as low as 10-15$M_\odot$, whose winds become sufficiently strong at an effective temperature of 21,000 K to cause ``pseudo-photospheres" to form, mimicking (at least) the behavior of many high luminosity LBVs.  (The 21,000 K effective temperature corresponds to the ``bistability jump", where a small drop in effective temperature results in a decrease in the ionization of the wind, leading to a smaller, but much denser, stellar wind.  See Pauldrach \& Puls 1990 and Lamers et al.\ 1995.)   
But, much of this is conjecture.  

What {\it is} clear is that due to the length of time between major outbursts, it is very tough to get a good handle on the number of LBVs.  The spectra of ``hot" LBVs show emission of [FeII], He I, and the Balmer lines, while the latter two
series usually display P Cygni profiles (see, for example, Figure 10 in Massey et al.\ 2007a).  Their spectra can instead resemble those of P Cygni itself, with strong Balmer and He I P Cygni emission but no forbidden Fe.  During their cool phase they can develop
a ``pseudo-photosphere", and their spectra resemble that of a late G- or F-type supergiant with strong Balmer emission.   The
``hot" phase is often described as their ``quiescence" stage, and their ``cool" phase as an ``outburst" stage, but an accidental spectrum of
S Dor itself in 1999 revealed the coolest spectrum in 50 years of spectroscopic monitoring; the F-type ``pseudo-photosphere" was
not accompanied by any major change in its photometry (Massey 2000).   Of course, that may require several hundred years (see discussion
in Massey 2006).  We will discuss this further in Section~\ref{Sec-LBVs}.

\subsubsection{Red and Yellow Supergiants}

Stars of somewhat lower mass ($<30 M_\odot$?) will evolve quickly to the red, passing first through a yellow supergiant phase.  It is in this stage that a massive star will be at its visually brightest, with the peak of the flux distribution being at visible wavelengths.  (Recall that stellar evolution of massive stars takes place at nearly constant bolometric luminosity, so as the star cools to temperatures 
similar to those of the Sun, their visual brightness will peak.)  The YSG phase is so short (a few thousand to a few tens of thousands
of years), that the phase has little evolutionary implications (i.e., in terms of mass loss), but as we shall see in Section~\ref{Sec-YSGs} this phase acts as a very sensitive probe of the accuracy of earlier evolutionary calculations.

At the coolest temperatures, these $< 30M_\odot$ stars are red supergiants, the physically largest stars.  Levesque (2010, 2012) has recently reviewed their physical properties.  For many years, their ``observed" locations in the H-R diagram (HRD) were much cooler and more luminous than
stellar evolutionary theory allowed, although few researchers seemed to realize this.  The basic problem was that the effective
temperatures were derived primarily from lunar occultations of red giants (see discussion in Massey \& Olsen  2003).  The use of
the new MARCS models (Gustafsson et al.\ 1975, Plez et al.\ 1992), which included sphericity and improved molecular opacities for TiO and other oxygen-rich molecules (Plez 2003, Gustafsson et al.\ 2003, Gustafsson et al.\ 2008), led to effective temperatures and luminosities that agreed well with the Geneva models over a wide range of metallicities (Levesque et al.\ 2005, 2006; Massey et al.\ 2009).  

The amount of mass lost during the RSG phase is a key factor in determining what happens next to the star; it is also one of factors most poorly constrained by observation and theory (Georgy et al.\ 2013).  High mass loss during this stage will shorten the RSG phase, and cause the star to evolve bluewards again in the HRD, going through a second YSG phase possibly becoming a Wolf-Rayet star or even an LBV (Groh et al.\ 2013).  Otherwise, RSGs are expected to end their lives as Type II supernovae. 

RSGs are ``smokey": dust condenses as the star loses mass.   Recently this led to the realization that RSGs suffer significant
circumstellar reddening (see, e.g., Massey et al.\ 2005a).  Little is understood about what drives this mass loss.  Massey et al.\ (2008)
quotes Stan Owocki (2007, private communication) as arguing that it doesn't take much to ``drive" a RSG wind. The escape velocity from a star is just 620 km s$^{-1} \times \sqrt{(M/R)}$.  Although  O stars have a M/R ratio that is of order unity,  RSGs do not: the ratio is much smaller, more like 0.02. So, the escape velocity is down by a factor of 7, less than 100 km s$^{-1}$.  Owocki argues that the mass loss of a hot star is set by conditions outside the stellar interior, i.e., opacity in the atmosphere and wind, that results in the radiatively-driven mass loss (Castor et al.\ 1975). For RSGs, the Òheavy liftingÓ has already been done by the stellar interior, as a significant fraction of the luminosity of the star has gone into making a bigger radius. ``It
is  kind of like walking with a nearly full glass of water vs a glass that is only 1\% full (O star)---even a small jiggle can 
lead to big changes in the mass loss for a RSG," Owocki concludes. 

\subsubsection{Wolf-Rayet Stars}

Finally, let us introduce the Wolf-Rayet stars.  These stars have strong emission-line spectra.  WR stars come in basically two flavors, WN-type WRs, where the spectrum is dominated by N and He, and WC-type WRs, where the spectrum is dominated by He, C, and O.  (A few WO-type WRs have been identified; these are basically WC stars with enhanced O lines.)
 In the ``Conti scenario" (Conti 1976), a massive star peels off its H-rich outer envelope through stellar winds on the 
 main-sequence (perhaps helped by the enhanced mass loss during the LBV phase).  Once the H-burning products (N and He) are revealed, the star
 is spectroscopically identified as a WN-type.  Further mass loss eventually peels off these layers, revealing the He-burning products (C and O).  
 
An outstanding question is why the mass-loss rates of WRs are so high; H-poor WRs have derived mass-loss rates about 10$\times$ higher than OB stars of the same luminosity.   The terminal wind velocities are similar, so it is as if the radiation pressure was somehow more efficient in these H-poor objects.  Puls et al.\ (2008) offers several possibilities, including the speculation that this is due to WR star winds being less ``leaky" because the higher core temperature and higher wind density leads to an ionization equilibrium that is stratified.  Alternatively, the
ionization in the outer winds may be shifted towards lower stages, leading to a more efficient acceleration.  Or it could be that WR winds get an extra kick due to the hot ion peak around 160,000 K deep in the WR atmosphere.  

What we do know is that the mass-loss rates on the main-sequence depend upon metallicity, as mentioned earlier, and that therefore it is easier for a massive star to become a WR at high metallicity than at low.  To put this more precisely,
a $20M_\odot$ might be able to ``peel down" all of the way to a WC star in M31 (where the metallicity is high, see Table 1), but could
only evolve as far as the WN stage in M33 (where the metallicity is lower, Table 1).  Or perhaps only stars of 60 $M_\odot$ and above
can become WRs in the SMC.  One expectation then is that the relative number of RSGs and WRs should be a sensitive function of
metallicity among the galaxies of the Local Group (Maeder et al.\ 1980), with relatively fewer WRs at lower metallicities.  In addition, one expects that the relative number of WC and WN stars should be a strong function of metallicity, with relatively fewer WCs at lower metallicities (Massey \& Johnson 1998, Meynet \& Maeder 2005, Neugent et al.\ 2012a).

\subsubsection{Supernovae (SNe)}

Massive stars are expected to end their lives as core-collapse SNe, enriching the interstellar material chemically, and providing a great deal of mechanical energy.  However, the only example we know of this directly in the Local Group is SN 1987A, where the progenitor star, Sk $-69^\circ$ 202, was known to be a $\sim 20M_\odot$ B-type supergiant.  At the time, evolutionary models did not include a ``blue loop" for such stars; 
they were supposed to become supernovae as RSGs.  

Core-collapse SNe are either hydrogen-rich type II SNe, or hydrogen-poor types Ib and Ic, where the latter
are also helium-poor.
The type II's are further classified depending upon their light curves (type II-P, whose light curves show a plateau, and II-L, where the light curve shows a magnitude-linear decline in time) or spectral features (IIn, where the spectrum shows narrow emission lines, and IIb, where the spectrum changes to become like that of a Ib).

The lack of hydrogen features in the spectra of types Ib and Ic  suggest that the progenitors have been been stripped of their outer hydrogen envelopes.  Two prevailing theories for how this stripping occurs are mass-loss through strong stellar winds or through binary interaction.  The former naturally associates the Ib/c's with Wolf-Rayet stars, but  there are no cases where an actual progenitor of a Ib or Ic has been identified.
(Shara et al.\ 2013 argue that the identification of Wolf-Rayet stars in nearby galaxies serves the additional purpose of identifying Ib/c progenitors, so that when one of these stars explodes we'll be prepared.)

All type II-P core-collapse SNe have long been assumed to come from RSGs,  although the classification
of SN 1987A is that of a (peculiar) II-P and we know it didn't.   Smartt et al.\ (2009) have identified numerous type II-P progenitors in nearby galaxies as RSGs, as well as summarizing those found by others.  They argue that type II-Ps come from a 
mass range from 8.5 to 16.5$M_\odot$.  Since RSGs have masses as large as 25$M_\odot$, this introduces
the ``red supergiant problem": why is the observed upper mass limit 16.5$M_\odot$.  Of course, this mass
limit is not directly observed; it is inferred by making a variety of assumptions not only about the IMF slope
in nearly galaxies, and it also depends upon certain reddening assumptions, etc.  

One of the most remarkable recent astronomical events has been the explosion (SN 2011dh) of
a YSG in M51 (distance 11 Mpc) as a Type IIb supernova (Maund et al.\ 2011).  Two other examples of YSG progenitors for SNe explosions exist; see Georgy et al.\ (2013) and references therein.  This fact poses some interesting, but not insurmountable, challenges
to single-star massive star evolutionary theory (e.g., Georgy et al.\ 2013), as 15$M_\odot$ stars (which is a typical mass for YSGs) have long been thought to end their lives in the RSG stage.  
But significant mass-loss during the RSG may delay the 
explosion (Bersten et al.\ 2012, Georgy et al.\ 2013).  Alternatively, these YSGs may have been binaries, and
binary evolution may have played a role in the formation of these SNe.

Of similar import has been the explosion of an 
LBV in NGC 7259 (distance 25 Mpc, far outside the Local Group),  
a Type IIn core-collapse supernova (SN); see Mauerhan et al.\ (2013) and references therein.
The star, SN 2009ip, was a transient which had recent eruptions. (In fact, the
first such eruption was mistaken for a SN explosion.)  As of this writing, the result is still somewhat controversial
(see, e.g., Pastorello et al.\ 2013), but if the discovery is confirmed, it will be remarkable in that for the first time we knew for certain what the progenitor was of a type IIn explosion, and as a demonstration that an LBV could explode as a SN.

\section{The Massive Star Content of the Local Group}

\subsection{Counting the OB Stars: Harder than it Looks}
\label{Sec-MS}

Obviously most of the massive stars in a galaxy will be of O- and B-type simply because these are the spectroscopic types associated with the main-sequence, and stars spend $\sim$90\% of their lives on the main sequence.  In addition, unlike the situation we will describe for yellow supergiants, and, to a lesser extent for red supergiants, there is little or none of the foreground contamination that
makes it difficult to separate bona-fide extra-galactic O and B stars from foreground stars.  Basically all blue stars in the field of a Local Group galaxy belong to that galaxy.

Yet, characterizing these massive stars is difficult.   We need both the bolometric luminosity and effective temperature of a star to place it on the HRD; at the very least we need the bolometric luminosity to even guess the mass.  And here we are hamstrung by physics: on the main-sequence these massive stars are so hot that the optical is way down on the tail of the Rayleigh-Jeans distribution.  The problem is humorously sketched out in Figure 2 of Conti (1986).   

\begin{figure}
\begin{centering}
\includegraphics[width=10cm]{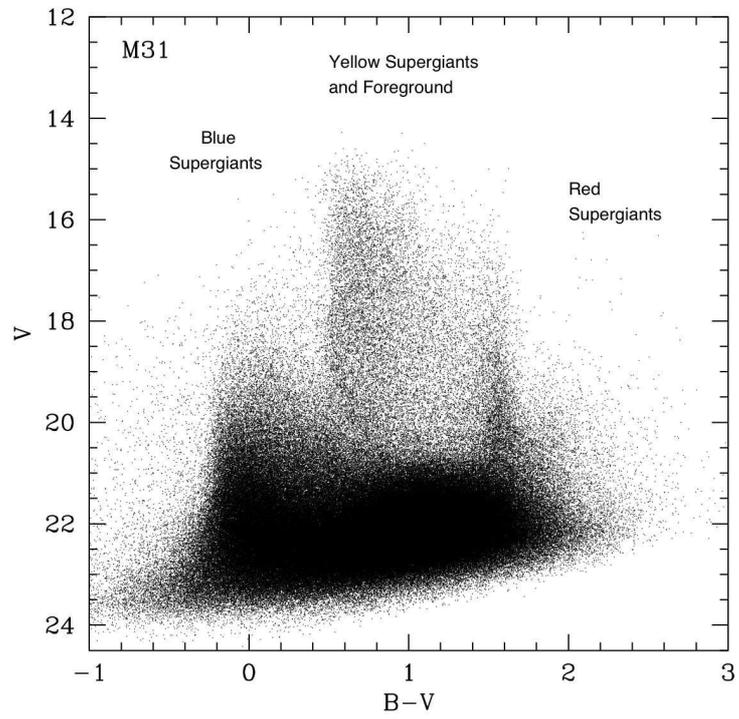}
\includegraphics[width=10cm]{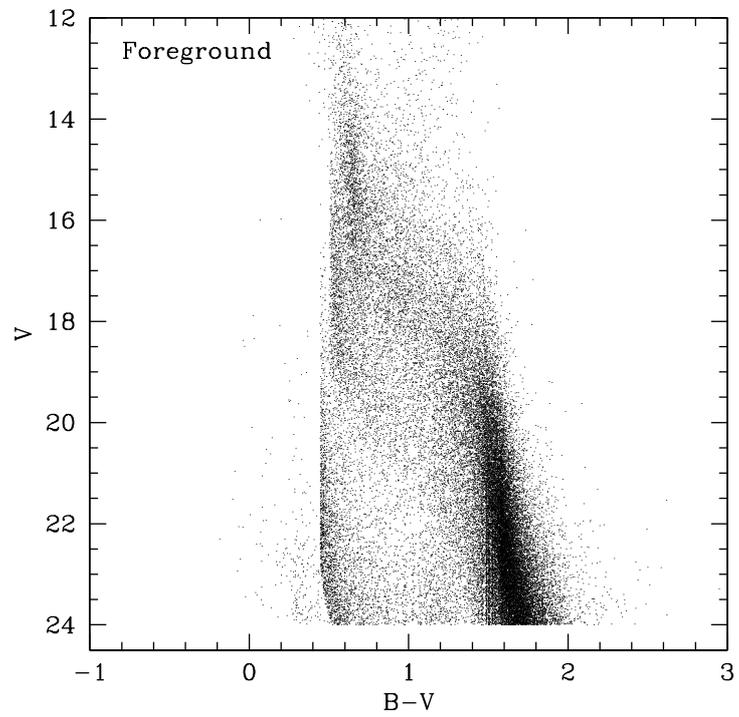}
\caption{\label{fig:CMD} Color-magnitude of M31.  The upper diagram shows the observed data, while the lower one shows the expected contribution from foreground stars.  From Drout et al.\ (2009) and used with permission.}
\end{centering}
\end{figure}

Good photometry is essential to the process, and for that reason the Local Group Galaxies Surveys (LGGS) was undertaken in the early 2000s, designed to obtain {\it UBVRI} photometry to 1-2\% or better for a 20$M_\odot$ star (Massey et al.\ 2006, 2007a, 2007b).  This is complicated by the fact that O-type stars are by no means the brightest stars in a galaxy.  They are the most bolometrically luminous, but they are several magnitudes fainter than, say, B or A supergiants of much lesser mass.  Typical $V$ magnitudes for a 20$M_\odot$ star on the zero-age main-sequence (ZAMS) run from 21.3 in M31, say, to 23.2 in IC 10, the latter due to its high reddening; in contrast, such stars are found about 6 magnitudes brighter in the LMC than in M31.   The realm of the OB stars in the color-magnitude diagram of M31 is shown in Figure~\ref{fig:CMD}. 

Such photometry allows us to determine the absolute visual magnitude if the extinction is known, or if it can be estimated using
a combination of a reddening-free color index plus the observed colors (see, e.g.,  Massey et al.\ 1989).  However, while necessary, photometry is not sufficient, not if we want to know the approximate mass of the star.   The reason is simple: the conversion from absolute visual magnitude to bolometric magnitude is a very steep function of the effective temperature, and the optical colors are highly insensitive to the effective temperature.  Massey (1998a) discusses this in detail, demonstrating that the Johnson $Q$ parameter changes by only 0.03 mag per 1 mag change in the bolometric correction.  This 1 mag change in the bolometric correction corresponds to a change of 0.2 dex in the $\log$ of the mass, assuming a mass-luminosity relationship of $L\sim m^{2.0}$.  In other words, even the best optical photometry would lead to a {\it minimum} uncertainty in the mass of 0.2~dex, the difference (roughly) between a 30$M_\odot$ star and a 50$M_\odot$ star.  In practice, 0.03~mag accuracy in $Q$ is very difficult to achieve (requiring 1\% photometry in both $U-B$ and $B-V$),
and a more realistic error is probably 0.05~mag in $Q$, corresponding to an uncertainty of 0.3~dex in the mass, or the difference
between a 30$M_\odot$ star and one of 65$M_\odot$.   Massey (1998a) goes on to demonstrate that even near-UV colors
(such as using {\it HST's} F170 filter, centered at 1700\AA) doesn't buy you more in terms of determining the effective temperature from photometry.

However, spectroscopy answers this neatly.  For O stars, the relative strengths of He I and He II are very temperature dependent. 
Modeling provides temperatures to about 1000~K or better, and at these temperatures that corresponds to an uncertainty in the bolometric correction of about 0.07~mag\footnote{For O stars, the bolometric correction goes as roughly $-6.90\log T_{\rm eff} +27.99$, so a change of 1,000 K at 40,000 K corresponds to 0.07~mag.}.  This translates to an uncertainty of only 0.015~dex in the mass, the difference between a 30$M_\odot$ star and one of 31$M_\odot$!  Even using spectral types to infer the temperatures leads to a similar improvement; an uncertainty of a single spectral type is a temperature difference of $\sim$2000 K (see, e.g., Table 9 in Massey et al.\ 2005b). 

In the last few years, we've seen an explosion in our knowledge of the massive star contents of the nearby universe.  The VLT-FLAMES survey of massive stars in the Magellanic Clouds (e.g., Evans et al.\ 2006, 2007;  Mokiem et al.\ 2006, 2007) and the
subsequent targeted study of the 30 Doradus region (e.g., Evans et al.\ 2011) have greatly improved our knowledge of the stellar content of our nearest  extragalactic neighbors.  These studies have not only produced spectral types, but also allowed detailed atmospheric analysis in many cases, e.g., Hunter et al.\ (2008). 

One disadvantage of such studies is in regions of high surface brightness nebulosity, such as the 30 Doradus region.  With fiber instruments (such as VLT-FLAMES) sky subtraction isn't local, unless time is spent using small offsets.  Nebula emission can contaminate the critical He I triplet lines, e.g., He I $\lambda 4471$.  Our group has started a program with Magellan to obtain high 
signal-to-noise long-slit optical spectroscopy of the early-type stars in NGC 346 and Lucke-Hodge 41.  The former is the brightest HII region in the SMC, while the latter is second in brightness and richness only to 30 Doradus itself.

A spectroscopic followup to the M31 and M33 LGGS requires large telescope aperture (given the faintness of the OB stars) and a multiplexing capability given the number of stars involved; this combination comes together in the 6.5-m MMT's Hectospec.  Preliminary results for $\sim 1700$ stars
were shown by Smart et al.\ (2012), and full results will be given by Massey et al.\ (2013).   Of course, this too suffers from nebular contamination in regions of strong nebulosity, and in those cases we are planning long-slit spectroscopy.

Garcia et al.\ (2009, 2010), Herrero et al.\ (2012), and Garcia \& Herrero (2013) have been doing similarly interesting work on the Local Group galaxy IC 1613, supplementing photometry with spectroscopy to study the young population.  (IC 1613 was the only star-forming Local Group galaxy to be left out of the LGGS.)  Their work has found a very interesting Of star that shows unexpected stellar wind characteristics among other discoveries (Herrero et al.\ 2012).  Bresolin et al.\ (2007) has also discussed spectroscopy for $\sim 50$ early-type stars in IC 1613.   Two O-type stars in WLM were discovered by Bresolin et al.\ (2006), along with $\sim$ 25 B-type supergiants.

\subsection{Luminous Blue Variables: Silent Quackers}
\label{Sec-LBVs}

Duncan (1922) and Hubble (1926) first discovered several very bright, irregular variables in M33 and M31, respectively. Among
these were Variable 1 in M33, and Variable 19 (AF And) in M31.  Hubble \& Sandage (1953) used photographic plates of M31 and M33 dating from 1920 to 1953 in order to study their properties, as well as to search for additional examples.  They found three
more in M33, which they designated Variables A, B, and C, but no new ones in M31.    They report on the characteristics of these five stars, noting that at their peaks they are among the brightest stars in these galaxies (reaching a photographic blue magnitude of 15th),
and showing variability of the order of multiple magnitudes, except for Var 19, where the amplitude was $\sim$ 1 mag.  In a footnote they likened these ``Hubble-Sandage variables" to that of S Doradus in the LMC.   Eventually the connection to Galactic stars like $\eta$ Car and P Cyg was made, and the term ``Luminous Blue Variable," or LBV,  was coined by Conti (1984).  Of the five original
Hubble-Sandage variables, Variable A is usually no longer listed as a ``true" LBV since its spectrum developed TiO bands; see Humphreys (1989).  At their most luminous, LBVs reach peak bolometric luminosities of about $10^6L_\odot$.

Identifying a complete sample of LBVs in Local Group galaxies is probably not possible.  Recall that
most Galactic LBVs are known because they have happened to have outbursts during historical times.  The last time the archetype LBV
P Cygni did
anything significant photometrically was in the 1600's.  
Massey (2006) has emphasized that if P Cyg were located in a nearby galaxy the star would not appear to be remarkable in any way unless we happened to take a spectrum of it.   

\begin{figure}
\begin{centering}
\includegraphics[width=13cm]{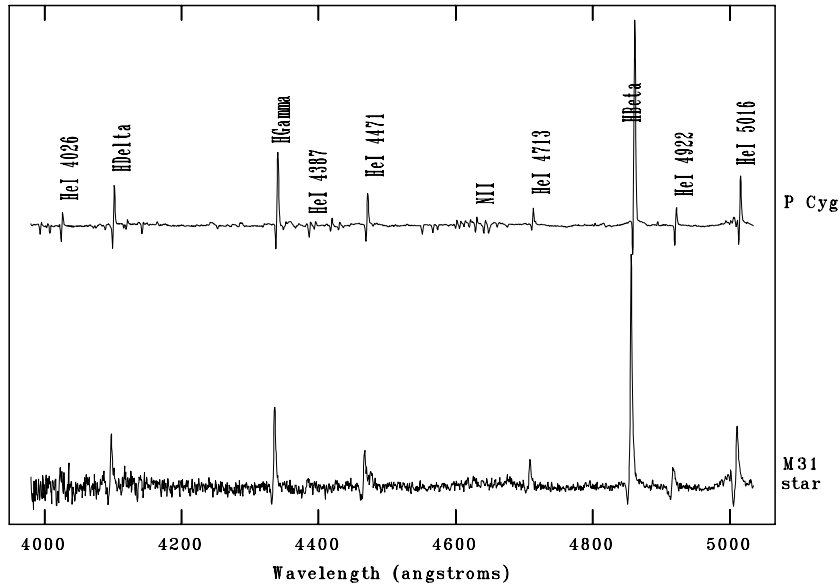}
\caption{\label{fig:PCyg}  A comparison between the spectrum of P Cyg and the M31 star  J004341.84+411112.0. 
From Massey (2006) and reproduced by permission.}
\end{centering}
\end{figure}

Indeed, Massey (2006) happened upon such a star in M31.  Its optical spectrum is indistinguishable from that of P Cyg; a comparison is shown in Figure~\ref{fig:PCyg}.   The helium and hydrogen lines show the characteristic P Cygni profiles, and even the NII lines are present, suggesting chemical enrichment of N at the surface. 
 
Massey et al.\ (2007a) undertook a survey of H$\alpha$ emission stars in the LGGS survey, finding many additional stars that were spectroscopically identical to known LBVs.   Are these stars true LBVs?  Bohannan (1997) argued that 
``A star should not be considered an LBV because its current spectroscopic character is similar to that of a known LBV. Remember what is said about ducks: it may look like a duck, walk like a duck, but it is not a duck until it quacks."   However, as Massey et al.\ (2007a) counter, ``In a raft of ducks, at any one time, some will be quacking and some will not. Momentary silence does not transform a duck into a goose, nor would it confuse most bird spotters."   One can further argue: if these stars are not LBVs,
then what are they?   In many cases their spectral signatures are unique.

One way that could resolve the issue is to see whether there is evidence of past ejecta events, and such a study is currently
underway using high spatial resolution {\it HST} spectroscopy in collaboration with Nathan Smith.  Such studies should reveal
if any of the LBV candidates in M31 and M33 have had past outbursts.  Future photometric monitoring is also planned.

\subsection{Yellow Supergiants: Looking at Stellar Evolution through a Magnifying Glass}
\label{Sec-YSGs}

As a massive star evolves to the red supergiant phase, it briefly becomes a yellow supergiant.  If the star also evolves back to the blue (expected for stars of about 30$M_\odot$), it will again pass through a YSG phase.  The length of this phase is measured not in million of years, but rather in thousands of years, about 0.1\% of the star's lifetime.  The brevity of this phase only adds to the usefulness of studying these objects.  
From a stellar evolutionary point of view, these stars act as ``a sort of a magnifying glass, revealing relentlessly the faults of calculations of earlier phases,"  as Kippenhahn \& Weigert (1990) put it.

Still, to capitalize on this phase, we must first identify a relatively complete sample.   The difficulty here is that there is a sea of foreground yellow dwarfs of the same color and magnitude range seen either against the Magellanic Clouds or the more distant
members of the Local Group.  For the red supergiants (discussed below in Section~\ref{Sec-RSGs}) we can separate foreground dwarfs
from extragalactic stars by the use of two-color indices.  But, no such photometric trick works for YSGs.

\begin{figure}
\begin{centering}
\includegraphics[width=8cm]{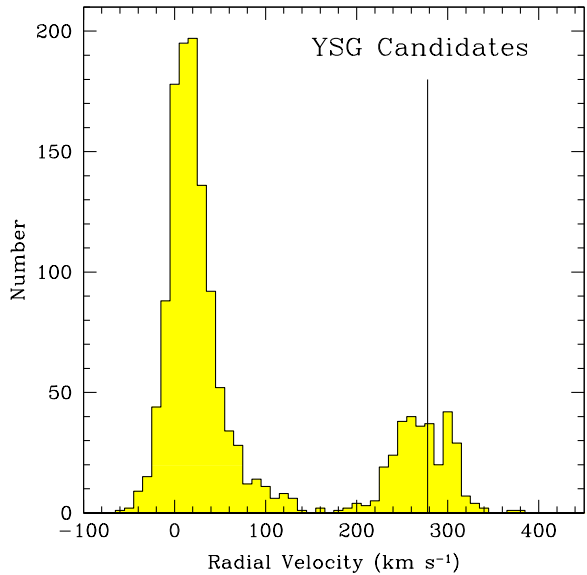}
\caption{\label{fig:YSGsRVs} The radial velocities of YSG candidates seen against the LMC. The vertical line at 278 km s$^{-1}$ shows the average radial velocity of the LMC.  
From Neugent et al.\ (2012b) and reproduced by permission.}
\end{centering}
\end{figure}

Instead, we must rely upon {\it radial velocities} to separate foreground contaminants from extragalactic YSGs.  Fortunately, this works extremely well.  In Figure~\ref{fig:YSGsRVs} we show the observed radial velocities of yellow stars seen against the LMC,  as found by Neugent et al.\ (2012b).
The stars are all in the right magnitude and color range to be YSGs in the LMC.  The vertical line at 278 km s$^{-1}$ shows the average
radial velocity of the LMC.  The errors on the individual velocities are small (of order 1 km s$^{-1}$); the dispersion about the systematic velocity of the LMC is due to that galaxy's rotation.  Nevertheless, it's clear that a very clean separation exists between the low velocity foreground yellow dwarfs and the LMC's YSGs.

How significant is this contamination?  For the LMC it's 80\% (Neugent et al.\ 2012b); i.e., only 20\% of the photometrically-chosen sample were found to be YSGs.  For the SMC (Neugent et al.\ 2010) it was similar, about 65\% were foreground.  The situation was even worse for M31 (Drout et al.\ 2009), where the contamination reached 96\%.

These studies have proven quite useful for evaluating the Geneva evolutionary models.  Drout et al.\ (2009) identified a sample of YSGs in M31 that was unbiased in luminosity.   Placing the stars on the HRD, they determined the relative number in each mass bin, and compared this to the predictions of the evolutionary models.  The number expected in the 20-25$M_\odot$ bin, compared to that in the 15-20$M_\odot$ mass bin will just be proportional to the relative lifetimes weighted by the initial mass function, assuming that the average star formation rate within a galaxy hasn't changed appreciably during the past few million years.  What they found was rather surprising.  On the one hand, the evolutionary models did a good job of predicting the locations of the YSGs in the HRD, in the sense
that they did not predict higher luminosity YSGs than those that were observed. However, in terms of the relative lifetimes, the models disagreed badly with the observation:  there were far fewer high luminosity YSGs than those predicted by the models.  
Neugent et al.\ (2010) extended this work to the SMC, with similar results.  Since the SMC is metal-poor, while M31 is metal-rich (see Table 1), this eliminated errors in the assumed mass-loss rates from the list of possible explanations.  

However, just as observational methods improve, so does theory.  Drout et al.\ (2012) and
Neugent et al.\ (2012b) extended these studies to M33 and the LMC.  However, by that time, new versions of the Geneva evolutionary models had become available (e.g., Ekstr\"{o}m et al.\ 2012) and both studies now found excellent agreement between the new models and the distribution of stars in the HRD.  As Drout et al.\ (2012) argued, the better agreement could not be due to a single cause, but 
due to a combination of improvements.  These include a better prescription for mass loss during the RSG phase, which affects the subsequent evolution from the red back to the blue side of the HRD.   The new models also adopt a new shear diffusion coefficient, and more realistic initial rotation velocities.   But, Drout et al.\ (2012) argue that even so, these changes by themselves could not explain the entire improvement.  Instead, the improved initial compositions and opacities must be important factors as well.  Regardless, while the older models did not do a good job predicting the relative numbers of YSGs as a function of luminosity, the newer models do an excellent job.

\subsection{Red Supergiants: Big and Sometimes Too Cool}
\label{Sec-RSGs}

The subject of red supergiants has been recently reviewed in this journal by Levesque (2010) and elsewhere by Levesque (2012), and the interested reader can find out much more about these fascinating objects in those two reviews.  Here we briefly summarize and complement these excellent papers.

\begin{figure}
\begin{centering}
\includegraphics[width=8cm]{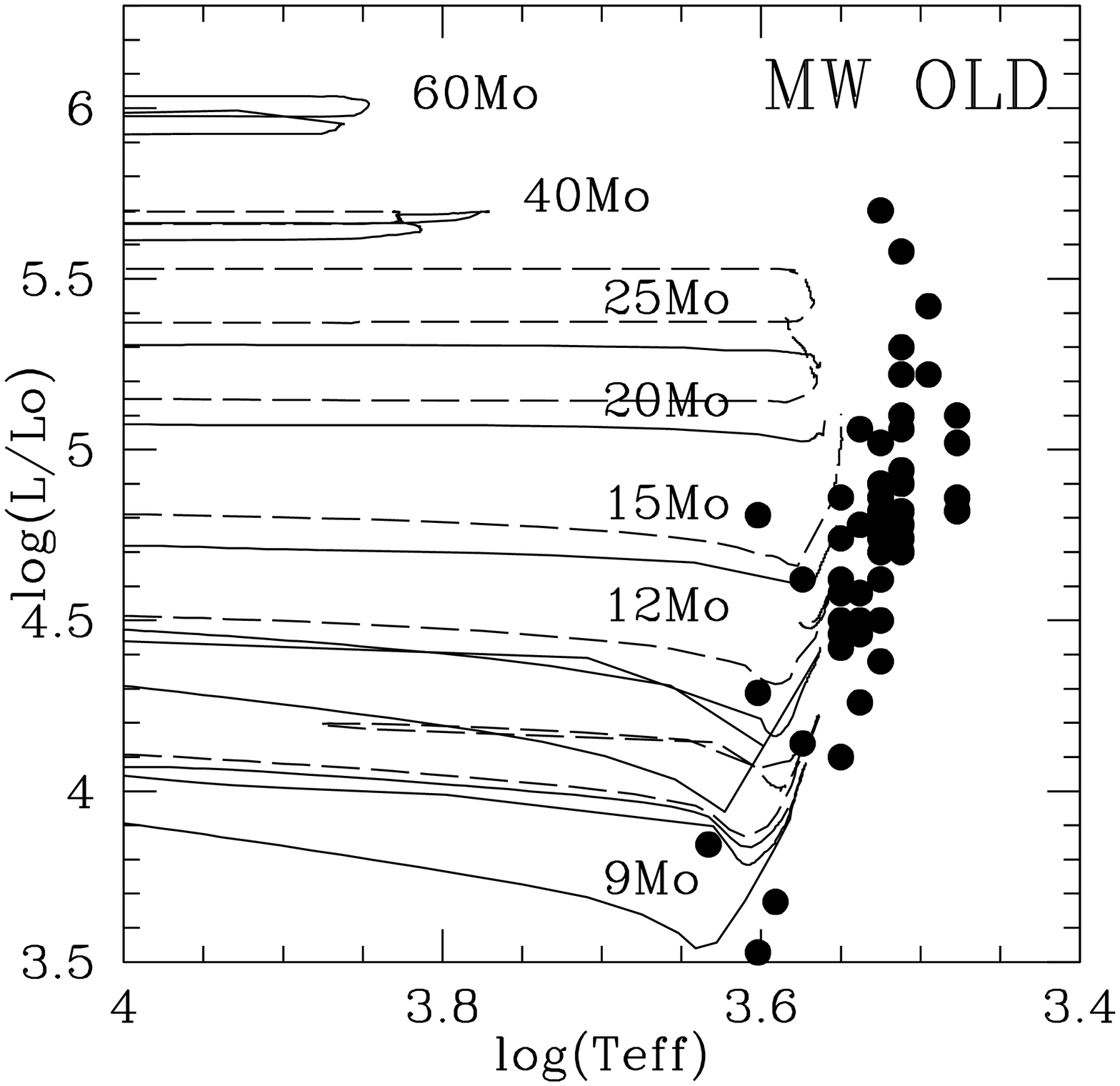}
\includegraphics[width=8cm]{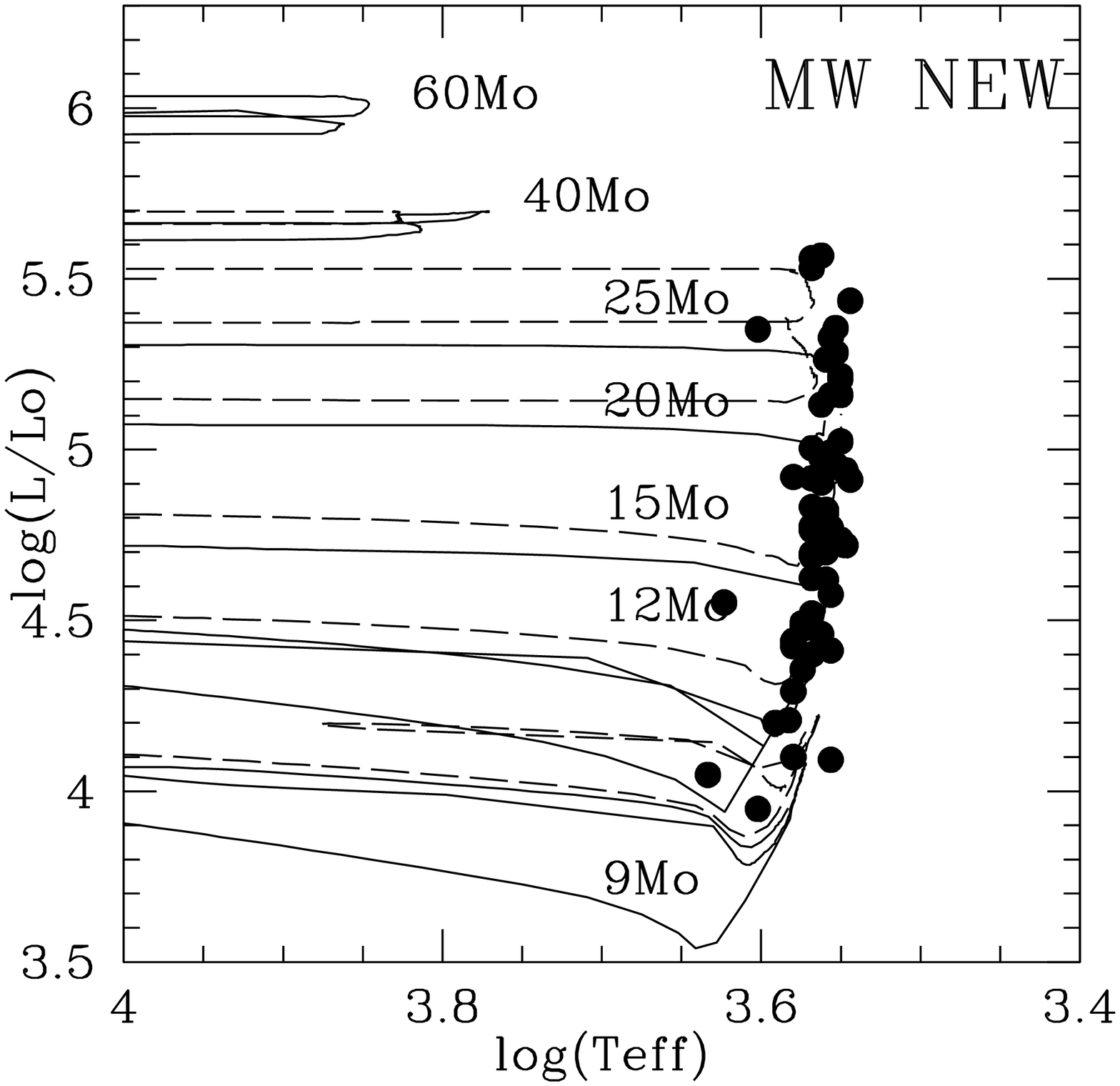}
\caption{\label{fig:RSGs} The effect of the new RSG effective temperature scale.  On the left is shown the poor agreement between the ``observed" location of RSGs in the HRD based on the old effective temperature scale and the evolutionary tracks.  On the right is shown the improvement from Levesque et al.\ (2005).  This figure is based upon one in Levesque et al.\ (2005) and  Levesque (2010).}
\end{centering}
\end{figure}

Most massive stars spend the majority of their He-burning lives as red supergiants,  and yet until relatively recently this phase
was poorly studied.  Massey \& Olsen (2003) were the first to note a significant discrepancy between the locations of RSGs and the
evolutionary tracks: if one used the standard spectral-type to effective temperature conversions, then RSGs were much cooler
than evolutionary tracks predicted.  Improved tracks became available, and yet the discrepency remained.  Using the newly available
atmospheric models, Levesque et al.\ (2005) determined a new effective temperature scale of Galactic stars, bringing into agreement for the first time the locations of RSGs in the HRD and the evolutionary tracks.  We show the improvement in Figure~\ref{fig:RSGs}.

\begin{figure}
\begin{centering}
\includegraphics[width=8cm]{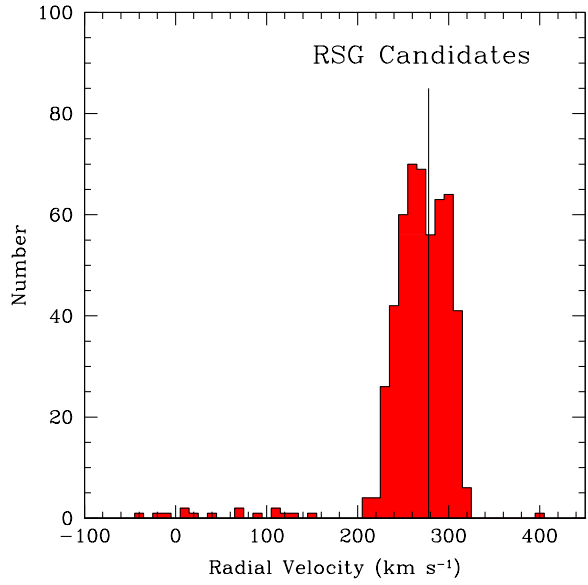}
\caption{\label{fig:RSGsRVs} The radial velocities of RSG candidates seen against the LMC. The vertical line at 278 km s$^{-1}$ shows the average radial velocity of the LMC.    For the RSGs in the LMC, foreground contamination is minimal, unlike the case for the YSGs (compare  to Figure~\ref{fig:YSGsRVs}).
From Neugent et al.\ (2012b) and reproduced by permission.}
\end{centering}
\end{figure}

Like the YSGs, identification of the RSGs among the galaxies of the Local Group is complicated by foreground contamination.
This turns out not to be much of a problem for RSGs in the Magellanic Clouds, as shown by Neugent et al.\ (2012b), and shown
here in Figure~\ref{fig:RSGsRVs}. (Compare this to Figure~\ref{fig:YSGsRVs}.)  But that is just because the RSGs in the Magellanic Clouds are bright ($V\sim 12-13$; see Neugent et al.\ 2012b and Levesque et al.\ 2006), and any foreground red dwarfs would have to 
be within about 50 pc of the sun to be of similar brightness.  The story is quite different for galaxies such as M31 and M33, where
the typical RSG will be 6 mags fainter, and red foreground dwarfs will be at the right brightness at distances of 800 pc, meaning that
the surface density of foreground objects will be 250$\times$ greater.

This problem is readily apparent if one compares the distribution of blue and red stars in M33; see, for example, Figures 21 and 22 of Humphreys \& Sandage (1980).  The blue stars are clumped, while the red stars show a far more uniform surface distribution.  The explanation for this cannot be one of age, as pointed out by Massey (1998b); rather, it must be that foreground contamination dominates.

\begin{figure}
\begin{centering}
\includegraphics[width=8cm]{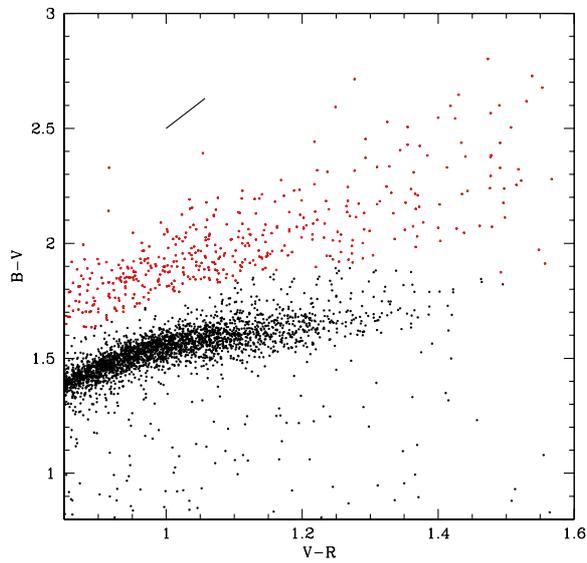}
\caption{\label{fig:M31RSG2COL} A two-color diagram of stars seen towards M31.  There are two distributions in $B-V$, with the
red points representing the (presumed) RSGs, and the black points representing the (presumed) foreground objects.  The reddening vector is indicated by the line near the top; its size is roughly that of the reddening seen towards M31 stars.  From Massey et al.\ (2009) and used with permission. }
\end{centering}
\end{figure}

\begin{figure}
\begin{centering}
\includegraphics[width=8cm]{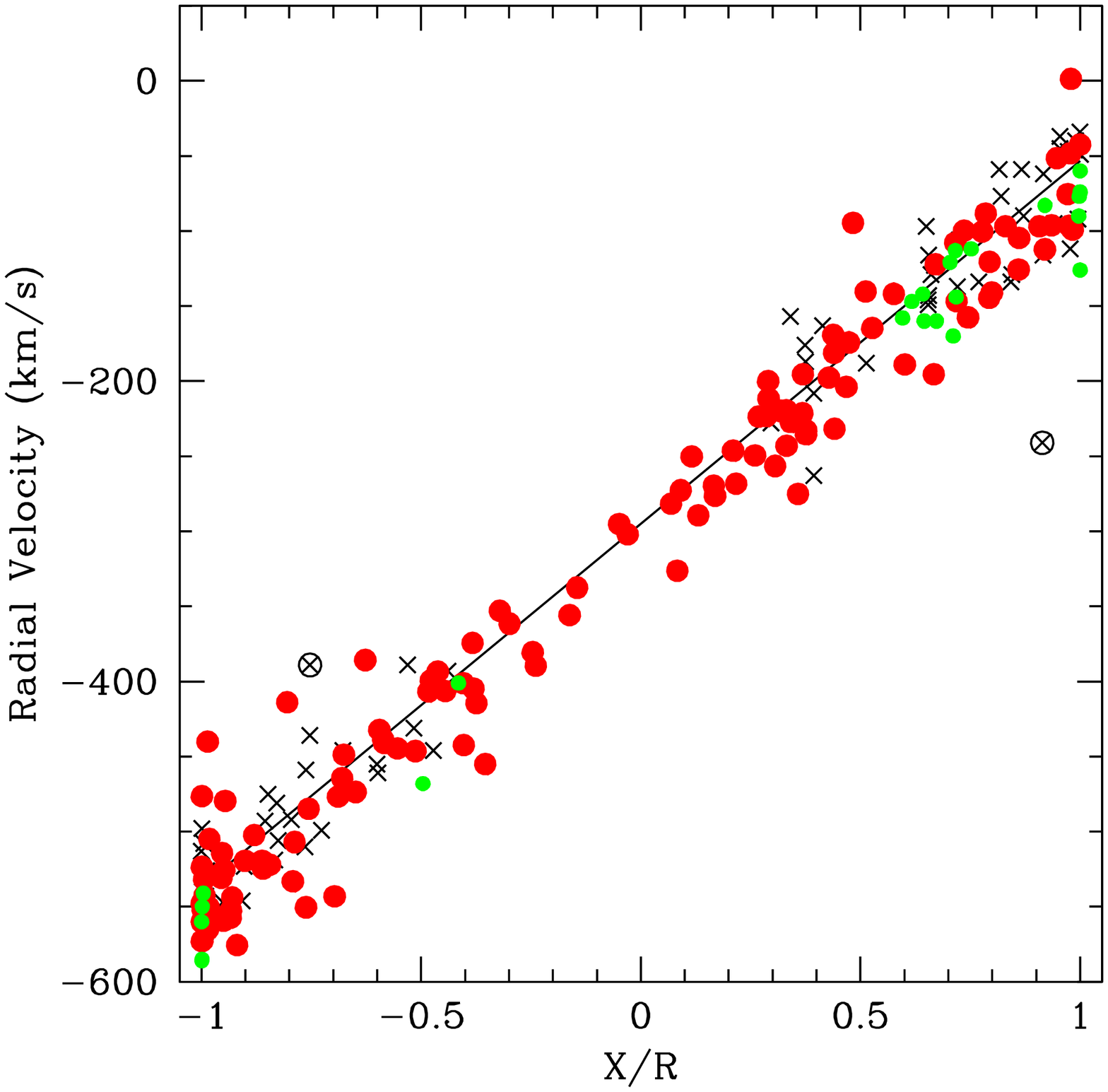}
\caption{\label{fig:M31RSGsRVs}  The radial velocities of the presumed RSGs (red points) from Figure~\ref{fig:M31RSG2COL} is shown superimposed on the rotation curve defined by the HII regions (crosses).  The green points show the stars previously confirmed
spectroscopically as RSGs by Massey (1998).   From Massey et al.\ (2009) and used with permission. }
\end{centering}
\end{figure}

Massey (1998b) found that it was relatively easy to separate extragalactic RSGs from foreground dwarfs by means of a $B-V$ vs $V-R$ two-color diagram.  At these low effective temperatures ($<$4000 K) $V-R$ is primarily a measure of effective temperature, while $B-V$ is dominated by surface gravity effects due to line-blanketing.   Such a diagram for red stars seen towards M31 is shown in Figure~\ref{fig:M31RSG2COL}.  The
dense band of (black) points are the presumed foreground stars, while the stars with the higher $B-V$ values (shown in red) are the RSG candidates.  Radial velocities demonstrated that the red points followed the rotation curve of M31, as shown in Figure~\ref{fig:M31RSGsRVs}.

To date, the RSG content of the LMC is relatively well established (Neugent et al.\ 2012b), as is that of M31 (Massey et al.\ 2009),
M33 (Drout et al.\ 2012), WLM and NGC 6822 (Levesque \& Massey 2012).  A comprehensive study of the RSG content of the SMC is still
lacking, although Levesque et al.\ (2006) successfully modeled a number of previously known SMC RSGs, determining effective temperatures and reddenings.  

One complication emphasized by Levesque (2010) is the potential confusion between RSGs and AGBs.  Normal AGBs can be separated from RSGs by using a luminosity cut off (Brunish et al.\ 1986) as done by Massey \& Olsen (2003).  Recently a class of 
``super"-AGBs has been postulated and discussed (Siess 2006, 2007, Poelarends et al.\ 2008); these are stars in a narrow mass range that ignite carbon off-center.  These stars could contaminate any sample of RSGs.

In general, our studies have shown that the coolest RSGs are warmer at low metallicities than at high metallicities; this is in
agreement with the finding of Elias et al.\ (1985) that the average spectral type of RSGs is earliest in the SMC, a bit later in the LMC,
and later still in the Milky Way, consistent with the progression in metallicity (see Table 1). This is a reflection of the shift in the
Hayashi limit  towards warmer temperatures with decreasing metallicity.  The Hayashi limit represents the
extreme effective temperature at which a star is still in hydrostatic equilibrium (Hayashi \& Hoshi 1972).  
However, a startling discovery by  Massey et al.\ (2007c) and Levesque et al.\ (2007) was that of a few of the most luminous RSGs and changed their spectral types quite significantly (e.g., K0 to M4) on the timescale of months.  At
the coolest these stars were much cooler than any of the other RSGs in their galaxies.   These stars show large photometric
variability, and changes in $A_V$, consistent with episodic dust production.  The nature of these objects is still unknown.

\subsection{The Wolf-Rayet Stars: Easy to Find Some, but Tough to Find the Rest}

Detecting complete samples of Wolf-Rayet stars sounds  like it should be easy. After all, their spectra are marked by strong emission lines, and should be readily found either by objective prism surveys or by interference filter imaging\footnote{An unconfirmed rumor has it that one Time Allocation Committee wag, debating one of the author's proposal for Kitt Peak 4-meter time to search for WRs in M31, claimed that he could find these using a 12-inch telescope in his backyard.}.  The problem, however, is that for a survey to be useful, it must be complete for both WCs and WNs if the relative ratios are going to be compared to that predicted by the evolutionary models.

\begin{figure}
\begin{centering}
\includegraphics[width=12cm]{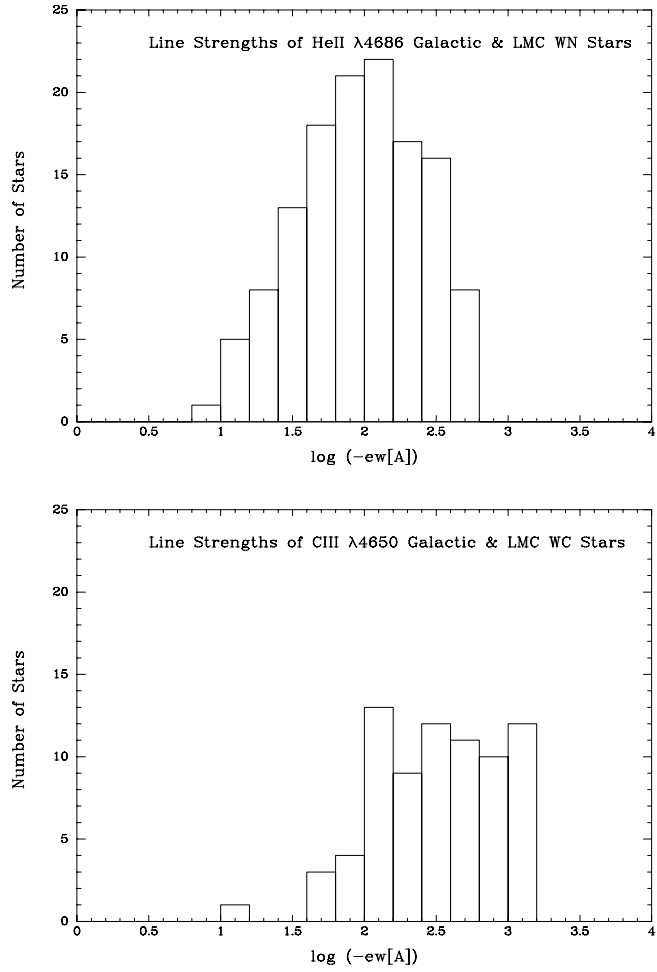}
\caption{\label{fig:EWs} The equivalent widths (ew) of the strongest optical lines in WN and WC stars are compared.  The data in this
figure comes from Conti \& Massey (1989), and the figure comes from Massey \& Johnson (1998).  Reproduced by permission.}
\end{centering}
\end{figure}

Surveys are intrinsically flux-limited; i.e., they are sensitive primarily to the emission line fluxes.  Let us consider the line strengths of
the strongest emission line in WC-type WRs, C III $\lambda 4650$, vs that of the strongest emission line in WN-type WRs, He II $\lambda 4686$.    The difficulty is thus illustrated in Figure~\ref{fig:EWs}, where we compare the equivalent widths (ews) of these strongest lines. We have used the equivalent widths as a proxy for line flux; we can do this (for the most part) as the absolute
visual magnitudes are quite similar between WC and (most) WN subtypes.  The exception are the WN7s, which are brighter
(see, e.g., Table 3-2 in Conti 1988).  The implications of this histogram is that WNs are harder to find, and thus one must be demonstrate that one is complete to sufficiently small line fluxes before claiming one has determined
the relative number of WC and WN stars.

Early efforts to detect WRs in the nearby galaxies of the Local Group have been nicely summarized by  Massey \& Conti (1983), Massey \& Johnson (1998),
Neugent \& Massey (2011) and Neugent et al.\ (2012a).   

In the SMC, an objective prism survey by Azzopardi \& Breysacher (1979a) used an interference filter to isolate the strong 
He II $\lambda 4686$ and/or C III $\lambda 4650$ lines, and found 4 new WRs, bringing the
total at the time to 8, with the previous 4 found by general spectroscopic studies (see Breysacher \& Westerlund 1978). A
ninth WR was found by  Morgan et al.\ (1991).  Massey \& Duffy (2001) undertook an on-band, off-band interference filter
imaging campaign with a wide-area CCD camera.  Photometry of 1.6 million stellar images helped identify a number of candidates,
including all of the known SMC WRs, at high significance levels. Follow-up spectroscopy then confirmed two new WNs, bringing the total to 11,  as well
as a number of Of-type stars, demonstrating that the survey was certainly sensitive enough to find even the weakest-lined
WNs.  However, shortly following this spectroscopy we accidentally discovered a 12th WR star in the SMC (Massey et al.\ 2003).  This
star had been too crowded to have been found in the Massey \& Duffy (2001) survey.   Of these 12 WRs, 11 are of WN-type and only 1 is of WC-type.  This low WC/WN ratio is consistent with the SMC's low metallicity.

For the LMC,  an objective prism survey by Azzopardi \& Breysacher (1979b) discovered 11 new WRs to add
to the already existing 80 that were known at the time (Fehrenbach et al.\ 1976), bringing the total to 91.
The relatively small increase probably was due to the limited coverage in the Azzopardi \& Breysacher (1979b) survey, with only a small fraction of the LMC surveyed. Since
that time, many other WRs have been found, mostly by accidental spectroscopy.  The latest catalog is that of  Breysacher et al.\ (1999), which lists 134 WRs.   Neugent et al.\ (2012c) discovered a rare WO star, and used the occasion to bring this list up to date by noting two stars that have been ``demoted" to Of stars,
and listing 6 other WRs since discovered by others.
This brings the number known in the
LMC to 139 stars, of which 107 are WN and 26 are WC or WO. 

The early discovery of $\sim$25 WRs in M33 by Wray \& Corso (1972)  really started 
the whole effort to discover WRs in Local Group galaxies beyond the Magellanic Clouds, and was responsible for the author's
growing interest in graduate school in the subject of studying massive stars in nearby galaxies.  Their work was followed by
Conti \& Massey (1981) and D'Odorico \& Rosa (1981), who found numerous WRs in M33's HII regions.   The first galaxy-wide systematic search (based on photographic on-band, off-band imaging) was described by Massey \& Conti (1983), who presented
spectral types and spectrophotometry for 79 WRs, about half of which had been previously known.  Armandroff \& Massey (1985)
and Massey \& Johnson (1998) provided CCD imaging in on- and off-band interference filters over a limited area to greatly improve the sensitivity and reveal mostly undiscovered WN-type WRs in M33.  (At the time of Massey \& Johnson (1998) the number of ``spectroscopically confirmed" WRs in M33 had grown to 141, although as subsequently shown by Neugent \& Massey 2011, a few
of these were bogus, i.e., quasars or Of-type stars.)  Finally, Neugent \& Massey (2011) conducted an imaging survey that covered all of M33 using the Kitt Peak 4-m telescope and Mosaic CCD camera, with follow-up spectroscopy with the 6.5-m MMT and the 300-fiber spectrograph Hectospec.  Their work found 55 new WRs, mostly of WN types, bringing the total known to 206, a number that they argue is complete to 5\%.  

\begin{figure}
\begin{centering}
\includegraphics[width=15cm]{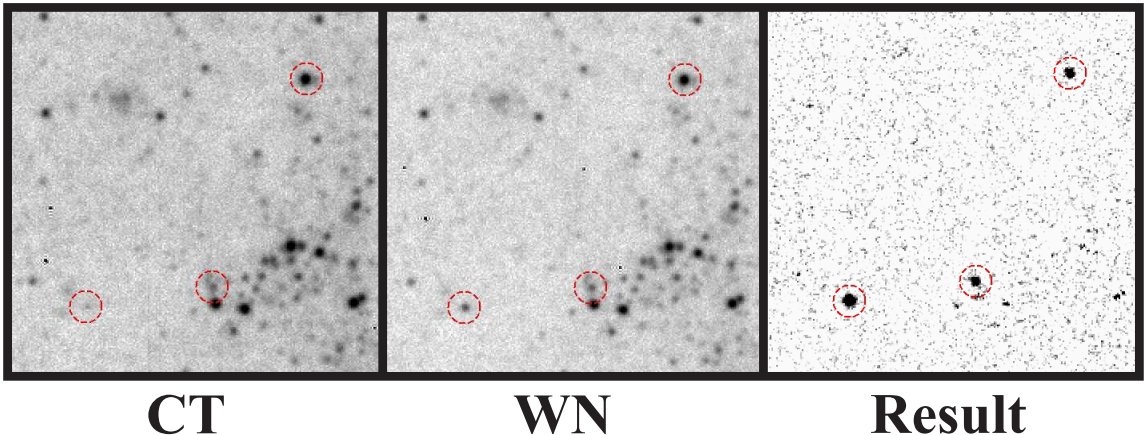}
\caption{\label{fig:imagesub} The power of the image subtraction technique is shown.  The figure on the left is the off-band continuum (``CT") image, the middle image is the on-band WR image (``WN"), and the difference image is shown on the right (``Result").
From Neugent \& Massey (2011) and reproduced by permission.}
\end{centering}
\end{figure}

Previous studies (e.g., Armandroff \& Massey 1985, Massey \& Johnson 1998, Massey \& Duffy 2001) replied upon doing photometry of all the stellar images and then looking for magnitude differences from the on-band to the continuum that were statistically significant.  This was not only time-consuming, but often failed because in some cases groups of stars would be photometered as 7 stars on one image but 6 on another: crowding matters.  It also tended produce a lot of false positives: Neugent \& Massey (2011) instead built on the image-subtraction techniques that had been developed over the years primarily to identify supernovae.  An example of how the technique works is shown in Figure~\ref{fig:imagesub}.

\begin{figure}
\includegraphics[width=8cm]{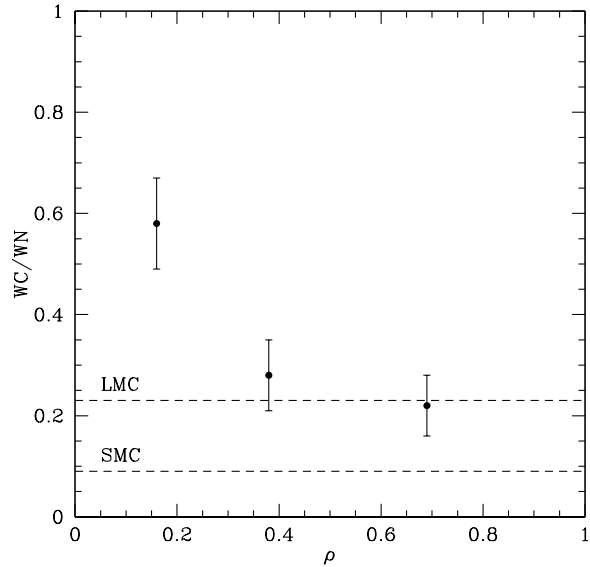}
\begin{centering}
\caption{\label{fig:m33wrgrad}  M33 WC/WN ratio vs galactocentric distance.  The ratio of the number of WC-type and WN-type WRs
is plotted against the galactocentric distance $\rho$ within the plane of M33.  A value of $\rho=1$ corresponds to 7.53 kpc.
From Neugent \& Massey (2011) and reproduced by permission.}
\end{centering}
\end{figure}

\begin{figure}
\begin{centering}
\includegraphics[width=8cm]{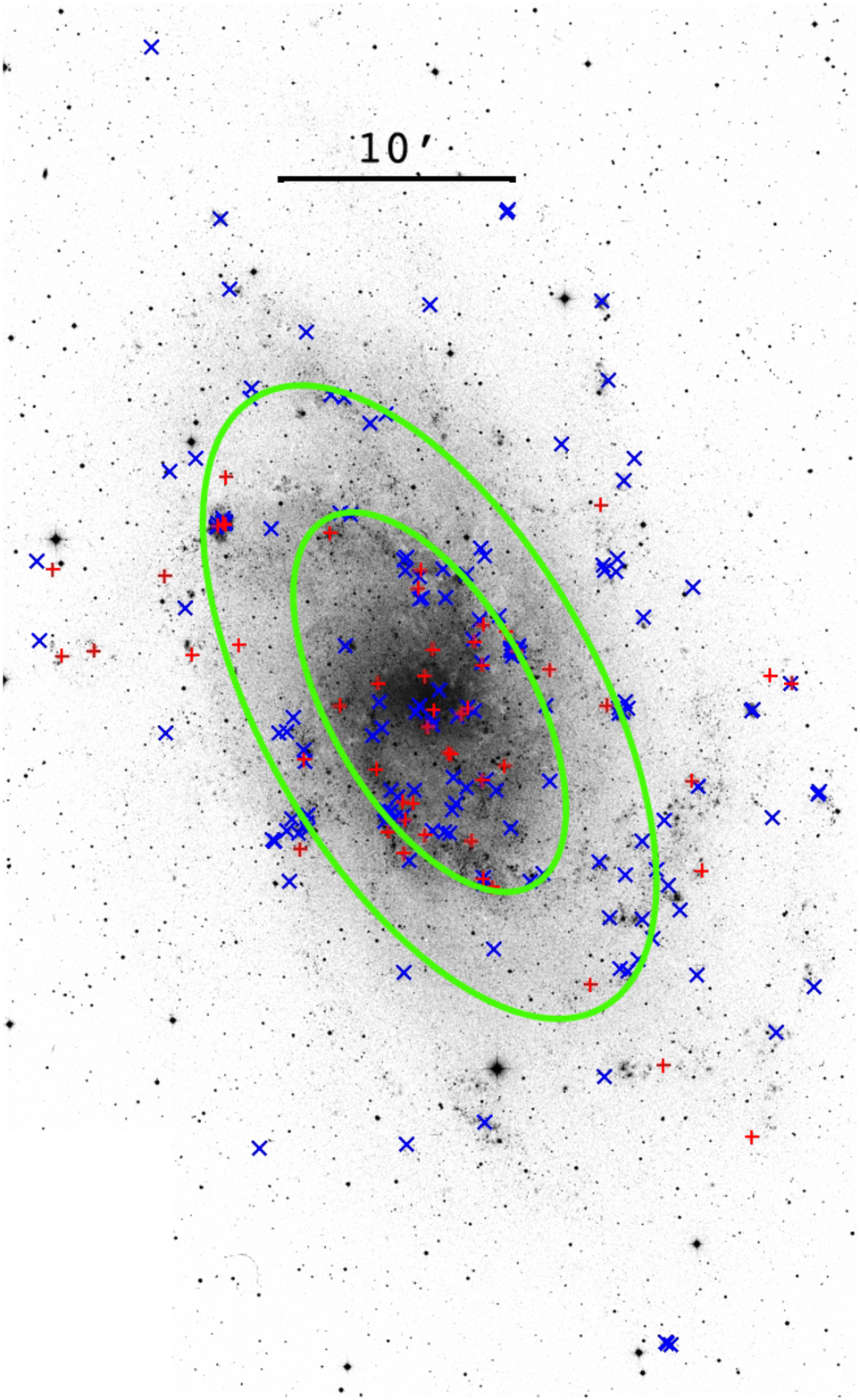}
\caption{\label{fig:m33wrs}  The distribution of WRs in M33.  The red +'s represent the WC-type WRs while the blue x's represent the WN-type WRs.  The two green ovals represent galactocentric distances of $\rho=0.25$ (1.9 kpc) and $\rho=0.50$ (3.8 kpc) within the plane of M33.  From Neugent \& Massey (2011) and reproduced by permission.}
\end{centering}
\end{figure}

One of the very attractive things about studying the WR content of M33 has always been the expectation that it has a metallicity gradient.  Would this be reflected in the relative population of WC and WN?  Even in the 1970s it was known that the SMC contained predominantly WNs while the LMC contained a better mixture of WC and WNs, and this was attributed to the fact the SMC had lower metallicity than the LMC.  Figure~\ref{fig:m33wrgrad} shows the results from Neugent \& Massey (2011)'s large sample,
where $\rho$ is the galactocentric distance within the plane of M33.  The distribution of M33's WRs in shown in Figure~\ref{fig:m33wrs}.

\begin{figure}
\begin{centering}
\includegraphics[width=8cm]{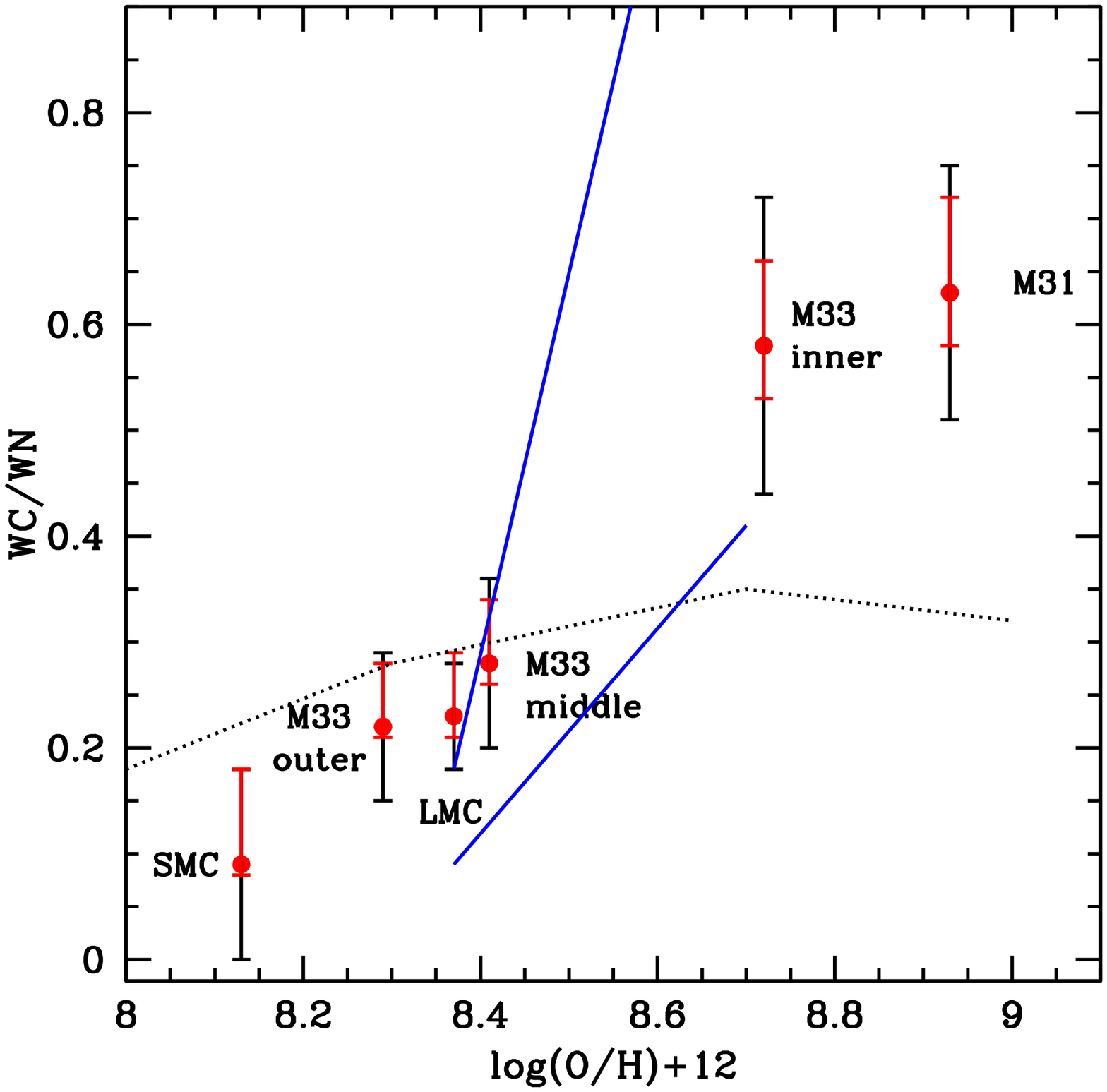}
\caption{\label{fig:modelfit}  The WC to WN ratio plotted against the oxygen abundance, which corresponds to 8.7 at z=0.014 (solar).
The red error bars indicate the measured uncertainties in the WC/WN ratio, while the black error bars show the
$\sqrt{N}$ uncertainties, which are more appropriate for comparing with the evolutionary model.  The two blue lines show the
prediction for the new Geneva models (the upper one being with no rotation, and the lower one being with
an initial rotation of 40\% of the break up speed), while the dashed line shows the prediction from the older Geneva models.  From Neugent et al.\ (2012a), and reproduced
by permission.}
\end{centering}
\end{figure}

In terms of WR surveys,  M31 has proven a very hard nut to crack, due to the galaxy's very large angular size.  At the same time, M31 is perhaps the most interesting,
as its metallicity is about two times solar (Zaritsky et al.\ 1994, Sanders et al.\ 2012; see discussion in Neugent et al.\ 2012a), providing
a key test of stellar evolutionary models.   Until recently, only photographic surveys had been carried out
galaxy-wide (Moffat \& Shara 1983, 1987), finding mostly WCs, presumably due to poor sensitivity. Massey et al.\ (1986) used much
deeper CCD imaging on a few selected OB associations, finding a better mixture of WNs and WCs.  The total number of WRs known
by the time of the analysis of Massey \& Johnson (1998) was 48, which gave a WC to WN ratio of 2.2.  Massey \& Johnson (1998) argued that
if the sample was just restricted to the regions that were complete, the WC to WN ratio would likely be much lower, about 0.93. Nevertheless this was much higher than the evolutionary models predicted (Meynet \& Maeder 2005). Neugent et al. (2012a) surveyed the entire galaxy, similar to the study of M33, bringing the total number of known WRs to 154, and yielding a WC to WN ratio of 0.67.  
This is still higher than the older Geneva models predict (i.e., Meynet \& Maeder 2005) but may be consistent with newer models once they become available at high metallicity.  (See Figure~\ref{fig:modelfit}.)

What about the irregular dwarfs beyond the Magellanic Clouds?  Westerlund et al.\ (1983) used a ``grism" to detect one WR (WN-type) star in NGC 6822; Armandroff \& Massey (1985) sampled most of this galaxy searching for WRs, and detected three other WNs that were eventually confirmed by spectroscopy (Massey et al.\ 1987).     A WO-type WR is known in IC 1613; it was discovered by D'Odorico \& Rosa (1982) as part of a survey of the ionizing stars of H II regions, and subsequently studied by Davidson \& Kinman (1982) and others (see, e.g., Massey et al.\ 1987, Kingsburgh \& Barlow 1995).  No additional WRs have been found.

The case of IC 10 deserves a special note.  Massey et al.\ (1992) found 22 WR candidates, of which 15 were confirmed by spectroscopy.   This was a spectacularly large number, given that IC 10 is about half of the size of the SMC (van den Bergh 2000),
and led to the galaxy's recognition as a starburst.  We now understand that the burst is being triggered by infalling gas from an extended cloud that is counter-rotating with respect to the galaxy itself, as shown by Wilcots \& Miller (1998).   Of the 15 spectroscopically confirmed IC 10 WRs listed by Massey \& Johnson (1998), 10 are of WC type, making the WC/WN ratio at that time be 2.  This was absurdly high given IC 10's relative poor metallicity (log O/H+12=8.2; see Table 1).  Even in the LMC, where log O/H+12=8.4 the
WC to WN ratio is 0.24, a factor of 8 lower.  This stood as a mystery until a much more sensitive survey was conducted by
Massey \& Holmes (2002).  This survey suggested that the total number of WR stars in IC 10 is even more remarkable than previously
thought, of order 100.   If so, this could bring the WC to WN ratio down to 0.3, about as expected, but would require a surface
density of WRs that is 20$\times$ greater than that of the LMC.   Only two of the new WR candidates had been confirmed
by Massey \& Holmes (2002) though.  Further follow-up is planned this coming year when Binospec is implemented on the MMT.

Beyond the Local Group, surveys for WRs have been conducted in NGC~300 (Schild \& Testor 1991, 1992; Breysacher et al.\ 1997;
Schild et al.\ 2003), M83 (Hadfield et al.\ 2005), NGC 1313 (Hadfield \& Crowther 2007), NGC 7793 (Bibby \& Crowther 2010), NGC 5068 (Bibby \& Crowther 2012), and M101 (Shara et al.\ 2013) among others.  The problem is, of course, that these galaxies are found at distances ranging
from 2.0~Mpc (NGC 300) to 7.0~Mpc (M83, M101), compared to, say, the distance to M33 (0.8~Mpc).  
So, the stars are going to be 1.9-4.6 ~mag
fainter, and crowding (although getting to be an issue in M33; see Neugent \& Massey 2011) a factor of 2.4-8.4$\times$ worse.
It would seem very difficult to identify an unbiased (much less complete) population of WRs in these galaxies if the goal is to compare
relative populations.

\section{Summary and Future Work}

We have briefly described the evolution and characteristics of massive stars, and described the usefulness of obtaining accurate
data amongst the galaxies of the Local Group.  Such information allows us to test models of how massive stars evolve at different
metallicities. And, we've seen that along the way there are many exciting new discoveries: a large population of LBV-like stars among Local Group galaxies, red supergiants that change their effective temperatures by large amount over the period of months, LBVs and yellow supergiants (admittedly in more distant galaxies) exploding as core-collapse supernovae in front of our very eyes, and more.

Many questions remain.  What are the true nature of the LBVs candidates?  How do the relative number of evolved massive stars of various kinds (LBVs, YSGs, RSGs) compared to the number of their progenitors?  
Why are some of the most luminous RSGs found at low metallicity highly variable in their spectral properties?
Why do the evolutionary models fail to predict as many WCs as observations show at high metallicity?
What is the true binary frequency of massive stars?

Work is continuing on characterizing the massive star populations in order to provide sensitive tests of massive star evolutionary models. At present, the most pressing and achievable goals are (1) to compare the binary frequency of WR stars in M31 and the center of M33 (where the models don't correctly predict the WC/WN ratios) with that in the outer portions of M33 (where the models do predict the right ratios), completing the WR survey for IC 10, characterizing the WR content of IC 10, identifying the nature of the LBV ``candidates" (based upon their spectral similarities to known LBVs) in M31 and M33, and better determining the numbers of RSGs in the SMC.  

Through such work, we gain not only a better understanding of massive stars, but provide the means for improving stellar evolutionary
models.  Such models are needed for population synthesis, allowing us to better understand how they
contribute to the ecology of more distant star-forming galaxies, and the production of supernovae of various types.  
Although studies of  exoplanets and high-redshift cosmology are often newsworthy, many exciting discoveries are awaiting to be made
in stellar astrophysics!

Ongoing collaboration with Maria Drout, Emily Levesque, Georges Meynet and especially Kathryn Neugent is gratefully 
acknowledged.  Peter Conti, Deidre Hunter, Emily Levesque, and Kathryn Neugent all  kindly made very useful comments on an early draft of this
manuscript.  This work was partially supported by the National Science Foundation through AST-1008020.

\end{document}